\documentclass[useAMS,usegraphicx,usenatbib]{mn2e}

\title[M33 Variability Survey]{Deep CFHT\thanks{Based on observations obtained with MegaPrime/MegaCam, a joint project of CFHT and CEA/DAPNIA, at the Canada-France-Hawaii Telescope (CFHT) which is operated by the National Research Council (NRC) of Canada, the Institut National des Science de l'Univers of the Centre National de la Recherche Scientifique (CNRS) of France, and the University of Hawaii.} Photometric Survey of the Entire M33 Galaxy I: Catalogue of 36000 Variable Point Sources}
\author[Hartman et al.]{J.~D.~Hartman,$^1$\thanks{E-mail: jhartman@cfa.harvard.edu~(JDH); dfb@astro.livjm.ac.uk~(DB); kstanek@astronomy.ohio-state.edu~(KZS); beaulieu@iap.fr~(JPB); jka@camk.edu.pl~(JK); marquette@iap.fr~(JBM); Peter.Stetson@nrc.ca~(PBS)} D.~Bersier,$^2$ K.~Z.~Stanek,$^3$ J.-P.~Beaulieu,$^4$ J.~Kaluzny,$^5$\newauthor
J.-B.~Marquette$^4$ and P.~B.~Stetson$^6$ \\
$^1$Harvard-Smithsonian Center for Astrophysics, 60 Garden St., Cambridge, MA~02138, USA\\
$^2$Astrophysics Research Institute, Liverpool John Moores University, Twelve Quays House, Egerton Wharf, Birkenhead, CH41~1LD, UK\\
$^3$Department of Astronomy, The Ohio State University, Columbus, OH~43210, USA\\
$^4$Institut d'Astrophysique de Paris, UMR70951~CNRS, Universit\'{e} Pierre \& Marie Curie, 98 bis boulevard Arago, 75014~Paris, France\\
$^5$Nicolaus Copernicus Astronomical Center, ul.~Bartycka 18, 00-716 Warszawa, Poland\\
$^6$Dominion Astrophysical Observatory, Herzberg Institute of Astrophysics, National Research Council, 5071 West Saanich Road,\\ Victoria, BC V9E 2E7, Canada\\}
\begin{document}
\maketitle

\begin{abstract}
We have conducted a variability survey of the local group galaxy M33 using $g^\prime$, $r^\prime$, and $i^\prime$ observations from 27 nights spanning 17 months made with the MegaPrime/MegaCam instrument on the 3.6 m CFHT telescope. We identify more than 36000 variable sources with $g^{\prime},r^{\prime},i^{\prime} \la 24$ out of approximately 2 million point sources in a one square degree field of view. This increases the number of known variables in this galaxy by more than a factor of 20. In this paper we provide a brief description of the data and a general overview of the variable star population which includes more than 800 candidate variable blue and red supergiant stars, more than 2000 Cepheids, and more than 19000 long period variable AGB and RGB stars.
\end{abstract}


\section{Introduction}

The study of variable stars in local group galaxies has played an important role in our understanding of the universe. Traditionally variable stars have served as one of the prime tools for determining the local distance scale. The discovery by Leavitt (1908; Leavitt and Pickering~1912) of the period-luminosity relation for Cepheids in the Magellanic clouds and the use of this relation by Hubble (1925, 1926, 1929) on Cepheids in NGC 6822, M33 and M31 enabled the modern view that these are all galaxies located outside the Milky Way.  

In recent years large-scale photometric variability surveys have revolutionized the study of variable stars. Dark matter searches such as EROS \citep{aub93}, MACHO \citep{alc97}, MOA \citep{nod02} and OGLE \citep{uda97} have released catalogues of more than two hundred thousand variables in the Galactic Bulge and more than sixty thousand variables in the Magellanic Clouds \citep{woz02,zeb01}. These surveys have shown that completely new science can be done with very large samples of variable stars. For example, a few of the results based on these catalogues include the discovery of multiple period-luminosity relations for Mira-like variables \citep{woo99}, the discovery of Beat Cepheids in the LMC \citep{alc95} and the discovery of a period-luminosity relation for ellipsoidal variations observed in LMC red giants \citep{sos04}.

While the Magellanic Clouds have been surveyed extensively by the microlensing searches, larger and more distant galaxies such as M31 and M33 have only been targeted with CCDs by relatively narrow-field variability surveys. With the advent of wide-field mosaic imagers it has become possible to perform a complete survey of these galaxies without the use of a dedicated telescope. Motivated by this opportunity, we have conducted a variability survey of the nearby galaxy M33 (01:33:51.00 +30:39:36.7 J2000) using the MegaPrime/MegaCam instrument on the 3.6 m Canada-France-Hawaii-Telescope (CFHT) atop Mauna Kea. 

There have been several previous variability studies of M33. The first variable stars in this galaxy were found by \citet{dun22} and \citet{wol23}. \citet{hub26} then found Cepheids in M33, thereby showing that this nebula is actually a galaxy at a distance of 263 kpc. \citet{hub53} investigated a handful of the brightest variables in M33 and discovered the Hubble-Sandage class (now called Luminous Blue Variables). Following this work \citet{van75} conducted a survey to find the brightest variables in M33 discovering 38 new variables. The last major photographic variability survey was conducted by \citet{kin87}. Most recently the DIRECT project conducted the first CCD variability survey of M33 discovering more than a thousand variables (Macri et al.~2001; Mochejska et al.~2001a,~2001b) in the centre of this galaxy. M33 has also been the target of several single (or few) epoch photometric surveys, with the survey by Massey et al. (2006, hereafter M06) as the most recent example. Our present survey has two improvements for finding variables over previous surveys of M33. The first is that by using a 3.6 metre telescope we go deeper than any previous survey. The second is that we are able to cover a much wider field (essentially the entire galaxy) than previously possible for CCD surveys. As a result we have detected 36709 variable sources which represents an increase by more than a factor of 20 in the number of known variables in the field of M33.

In the following section we discuss the observations. In \S 3 we describe the data reduction and we present the catalogue of variable sources in \S 4. We describe a few of the ensemble properties of these variables in \S 5 and conclude with a few summary remarks in \S 6.

\section{Observations}

The data were obtained using the Queue Service Observing mode at the CFHT on 27 separate nights between August 2003, and January 2005. We used the $g^{\prime}$, $r^{\prime}$ and $i^{\prime}$ Sloan filters, obtaining 36 images centred on the galaxy with each filter. The seeing varied from $0.6-1\arcsec$ in $i^{\prime}$. For the 2003 observations we used an exposure time of 530 s for $g^{\prime}$, and 660 s for $r^{\prime}$ and $i^{\prime}$. In the 2004 season we switched to 480 s for $g^{\prime}$ and 600 s for $r^{\prime}$ and $i^{\prime}$. Table~\ref{tab:obs} lists the dates of observations, the number of images obtained in each filter on these dates, and the median seeing for each filter.

\begin{table}
\begin{minipage}{84mm}
\caption{Observations of M33}
\begin{tabular}{@{}lrrrrrrr@{}}
\hline
 Date  &  $g^{\prime}$\footnote{Number of images in this filter aquired on this date.} & $r^{\prime}$ & $i^{\prime}$ & ${\rm FWHM}_{g^{\prime}}$\footnote{Median FWHM in this filter} & ${\rm FWHM}_{r^{\prime}}$ & ${\rm FWHM}_{i^{\prime}}$\\
 \hline
2003-08-22 & 1 & 1 & 1 & 0.72 & 0.70 & 0.61 \\
2003-08-24 & 2 & 2 & 2 & 0.93 & 0.91 & 0.78 \\
2003-08-30 & 1 & 1 & 1 & 0.95 & 0.82 & 0.69 \\
2003-08-31 & 2 & 2 & 2 & 0.92 & 0.90 & 0.72 \\
2003-09-03 & 1 & 1 & 1 & 0.81 & 0.75 & 0.64 \\
2003-09-21 & 1 & 1 & 1 & 0.95 & 0.84 & 0.71 \\
2003-09-23 & 1 & 1 & 1 & 0.81 & 0.79 & 0.76 \\
2003-09-26 & 1 & 1 & 1 & 0.76 & 0.67 & 0.68 \\
2003-09-27 & 1 & 1 & 1 & 0.89 & 0.85 & 1.03 \\
2003-10-19 & 4 & 4 & 4 & 1.00 & 0.92 & 0.84 \\
2003-10-28 & 1 & 1 & 1 & 0.86 & 0.85 & 0.77 \\
2003-11-20 & 1 & 1 & 1 & 1.03 & 1.14 & 0.97 \\
2003-12-18 & 1 & 1 & 1 & 0.95 & 1.06 & 0.87 \\
2004-08-23 & 1 & 1 & 1 & 0.95 & 0.99 & 0.84 \\
2004-09-07 & 1 & 1 & 1 & 0.87 & 0.85 & 0.72 \\
2004-09-12 & 1 & 1 & 1 & 1.06 & 0.79 & 0.61 \\
2004-09-13 & 1 & 1 & 1 & 1.27 & 1.20 & 0.98 \\
2004-10-08 & 1 & 2 & 3 & 0.89 & 0.99 & 0.82 \\
2004-10-09 & 1 & 1 & 1 & 0.76 & 0.75 & 0.83 \\
2004-10-18 & 1 & 1 & 1 & 1.19 & 1.20 & 0.95 \\
2004-11-03 & 1 & 1 & 1 & 0.99 & 0.82 & 0.69 \\
2004-11-07 & 3 & 1 & 3 & 0.98 & 1.00 & 0.83 \\
2004-11-08 & 1 & 1 & 1 & 0.98 & 0.89 & 0.84 \\
2004-11-13 & 1 & 1 & 1 & 1.21 & 0.96 & 0.78 \\
2004-11-19 & 1 & 1 & 1 & 0.88 & 0.85 & 0.67 \\
2004-12-06 & 1 & 1 & 1 & 0.97 & 0.83 & 0.75 \\
2005-01-16 & 1 & 1 & 1 & 1.20 & 1.17 & 0.99 \\
\hline
\end{tabular}
\label{tab:obs}
\end{minipage}
\end{table}

\begin{figure*}
\includegraphics[width=168mm]{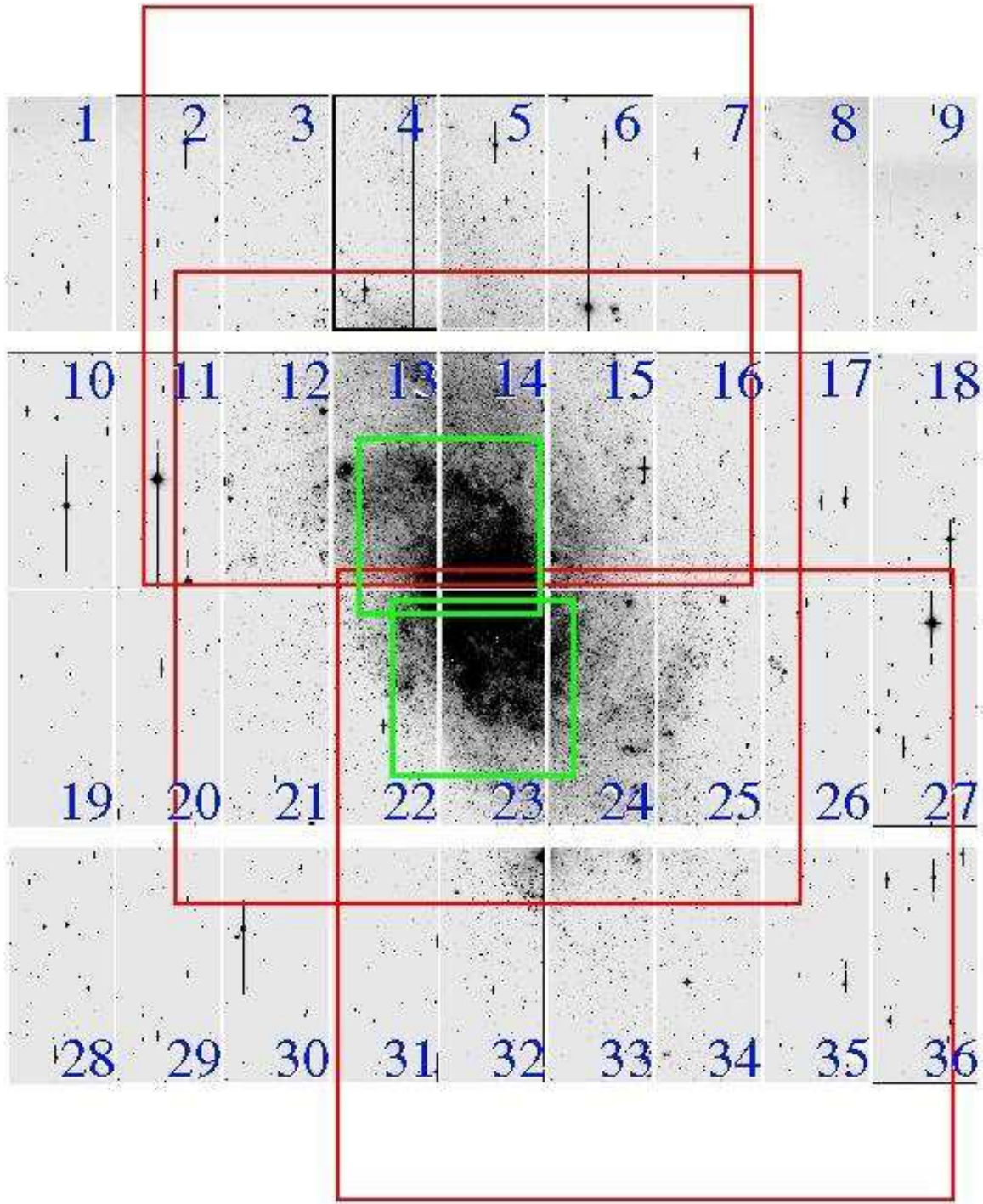}
\caption{A 36 CCD mosaic $g^{\prime}$ image of M33 obtained with the MegaPrime/MegaCam instrument on the 3.6 m. CFHT telescope. The field of view is 1 degree x 1 degree, north is up and east is to the left in this image. The smaller squares over the centre of the galaxy show the field of view of the DIRECT variability survey, while the larger squares show the coverage of the M06 photometric survey. The numbers for the chips are as used in the catalogue.}
\label{fov}
\end{figure*}

The MegaCam instrument is a wide-field mosaic imager consisting of 36 2048 x 4612 pixel CCDs. When used at the MegaPrime focus one obtains a 1 degree x 1 degree field of view with a sampling of $0\farcs187$ per pixel. An example mosaic $g^{\prime}$ image is shown in fig.~\ref{fov}, for comparison we also show the fields covered by DIRECT and the fields observed recently by M06. It is plain to see that with a single pointing it is possible to cover essentially the entire galaxy.

\section{Data Reduction}

The preliminary mosaic CCD calibrations are performed as part of the CFHT Queue Service Observing mode. The images are automatically processed through the Elixir pipeline which performs bias, dark, flat-field and fringe corrections, merges amplifiers, and provides an estimate of the photometric calibration to Sloan standards.

The data reduction procedure consists of two distinct steps: in the first step we use image subtraction to identify variable sources and obtain differential flux light curves, in the second step we perform point spread function (PSF) fitting to convert the differential flux light curves into magnitude light curves. The reductions were performed independently for each chip in each filter.

\begin{figure*}
\includegraphics[width=168mm]{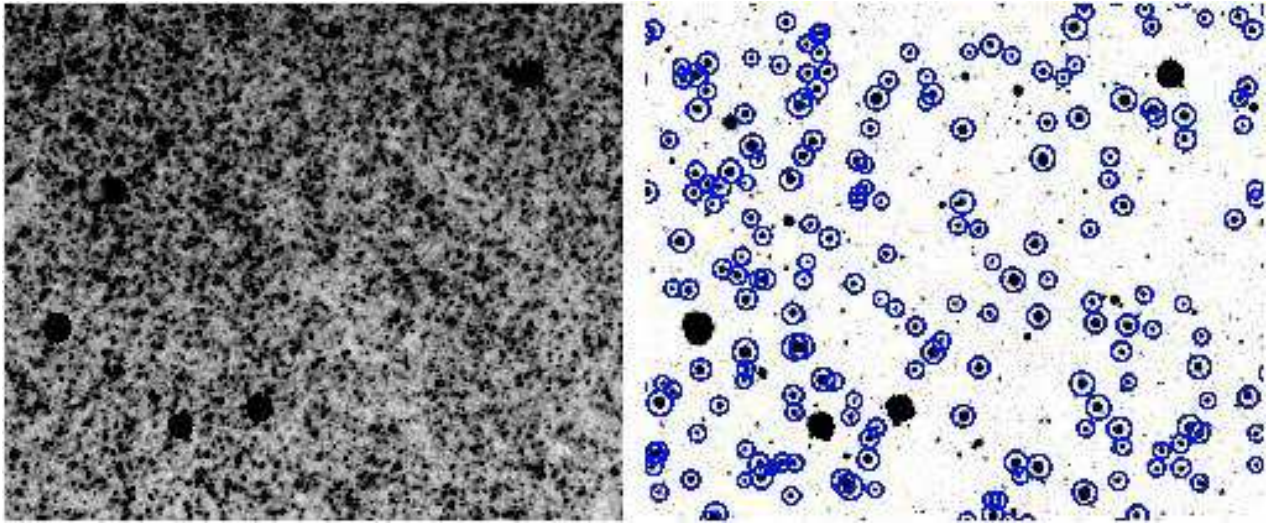}
\caption{Left: A $1\farcm5$x$1\farcm5$ portion of the $i^{\prime}$ reference image for Chip 14. Right: The corresponding ``var'' image from ISIS produced by co-adding in absolute value the residual images after rescaling by the square root of the original images. The circles show sources included in the catalogue of variables with the radius of the circle proportional to $\log(\chi^{2}_{N-1})$ of the light curve for that source. A few very bright regions on the ``var'' image and reference image are saturated stars that have been masked.}
\label{var}
\end{figure*}

To obtain light curves we use the image subtraction techniques due to Alard and Lupton (1998; Alard 2000) as implemented in the ISIS 2.1 package. This technique involves first registering the images to a master image. As a result we do not obtain light curves for any stars located on the chip gaps of the master image, we considered the simplicity of the standard ISIS scheme worth this sacrifice. We then combine a number of the best seeing images (we typically used 6 for each chip in each filter) into a master reference image. ISIS then solves for a transformation that matches the PSF, flux scale and background of the master reference and each image. The residuals between the transformed reference and each image are then co-added (in absolute value) and searched for significant residual point sources which are identified as variables. A portion of the result from co-adding the residual images for chip 14 (as labelled in fig.~\ref{fov}) in $i^{\prime}$ is shown in fig.~\ref{var} together with the corresponding portion of the reference image. Each of the residual images has been divided by the square root of its original image before co-adding their absolute values to create this image (this corresponds to the ``var'' image produced by the {\it findvar} routine in ISIS). As a result, the value of each pixel is roughly proportional to its significance of variability. We also plot the variable sources from our catalogue in the image as open circles with the radius of the circle proportional to $\log(\chi^{2}_{N-1})$ of the corresponding light curve where
\begin{equation}
\chi^{2}_{N-1} = \frac{1}{N-1}\sum_{i=1}^{N}\left ( \frac{m_{i}-\overline{m}}{\sigma_{i}} \right ) ^{2},
\label{eq:chisq}
\end{equation}
$m_{i}$ is the magnitude at observation $i$, $\overline{m}$ is the average magnitude of the light curve, and $\sigma_{i}$ is the magnitude uncertainty for observation $i$.

A number of the variable sources detected in this fashion are the results of artifacts in the data including subtle bad columns, slight jumps at amplifier boundaries, and diffraction spikes, bleed columns and reflections from particularly bright stars. We attempt to eliminate the latter by growing masks over bright regions connected to saturated pixels. This is not 100\% effective, and a few particularly pernicious regions are eliminated by hand. To eliminate artifacts due to bad columns/rows we look for and remove spikes in the histograms of variables vs. column or row number. In order to provide an unbiased estimate of the fraction of stars that are variable, we keep track of the eliminated regions and apply the same cuts to the point source lists determined below. We note that even with these cleaning procedures a number of artifacts may still slip into the catalogue. 

Once the variables have been identified we obtain their differential flux light curves by performing PSF fitting photometry on the residual images. We obtained light curves for all variable sources that passed the above cleaning procedures as well as a number of randomly selected non-variable sources in each chip that can be used to measure the photometric uncertainty and determine the significance of detection for the variables.

In order to convert the differential flux light curves from image subtraction into magnitudes we perform PSF fitting photometry on the master reference images using the DAOPHOT/ALLSTAR package (Stetson\ 1987, 1991). Due to the large amount of data we ran the procedure in batch mode, relying on the the automated ``pick'' routine in DAOPHOT to choose PSF stars. We perform the photometry in two passes, in the second pass we add in stars that are found on the image after subtracting off stars fitted in the first pass. For each chip we determine an optimal aperture correction radius to use in conjuction with ISIS that provides a minimum scatter among bright stars. We also determined an aperture correction at a fixed radius (13 pixels, or roughly 3 times the FWHM) to allow for a meaningful comparison of magnitudes from separate chips. We performed the latter correction directly on the model PSF to avoid contamination due to crowding. Note that the flux scale for each master reference chip in a given filter comes from the same image. 

After performing PSF photometry we match the variable sources detected by ISIS to the point source catalogue from DAOPHOT. We only include variable sources that match within 2 pixels ($0\farcs37$) to a point source. We use the DAOPHOT positions in measuring the light curves. As a final cleaning step we re-scale the magnitude uncertainties of all light curves on a given chip so that $\chi^{2}_{N-1}$ as defined in eq.~\ref{eq:chisq} for the constant light curves scatters about unity and does not correlate with magnitude. We then measure the standard deviation of $\chi^{2}_{N-1}$ for the constant light curves and require that a variable source must have a light curve with $\chi^{2}_{N-1}$ that is more than 5 standard deviations away from unity. Following the cleaning procedures we identified 7640 variables sources in $g^{\prime}$, 15770 in $r^{\prime}$, and 27961 in $i^{\prime}$.

The final catalogue consists of 36709 distinct variable sources. We provide $g^{\prime}$ light curves for 26342 of these sources, $r^{\prime}$ light curves for 30386 and $i^{\prime}$ light curves for 34320. For a given variable we provide a light curve in every filter in which the object was detected by DAOPHOT. To match the point source catalogues from separate filters we used the ISIS interpolation routine to transform the $g^{\prime}$ and $r^{\prime}$ image coordinates to $i^{\prime}$ coordinates and then matched the catalogues with a 2 pixel ($0\farcs37$) matching radius. There are 201 close pairs (two objects separated by less than $0\farcs3$) in the final catalogue, we note in the comments section of the catalogue if a given source is a member of a pair.

To provide approximate calibration to the Sloan standard filters we used the photometric calibration provided by the Elixir routine. We will provide a better calibration with the full point source catalogue in a future publication. In implementing the Elixir calibration we only use the zero-point terms and ignore the colour terms. We do this for simplicity in applying the future calibration. Please note that the magnitudes provided in this catalogue should at present be regarded as instrumental magnitudes, they may differ from the standard system by as much as a few tenths of a magnitude, depending upon the star's colour. We felt that the data are of sufficient value that it is worth releasing our variable stars catalogue, even though the photometric calibration is not definitive.

We used the point source catalogue from M06 to determine the astrometric solution for each chip. The astrometry for M06 is based on the USNO-B1.0 catalogue \citep{mon03}. For the few chips that do not overlap with M06 we match directly to the USNO-B1.0 catalogue. The errors in this match were typically less than $0\farcs2$. For stars with $R < 20$ the median matching error to M06 is $0\farcs13$. 

\section{Catalogue of Variable Point Sources}

The catalogue of variable point sources and the $g^{\prime}$, $r^{\prime}$ and $i^{\prime}$ instrumental magnitude light curves are available on the web\footnote{http://www.astro.livjm.ac.uk/\~{}dfb/M33/}, the full catalogue will also be included in the electronic version of the journal, in tables~\ref{tab:varcat1}-\ref{tab:varcat3} we show a few rows to illustrate the available data. We split the catalogue here into four separate tables so that they each fit on a single page. 

\begin{table*}
\centering
\begin{minipage}{168mm}
\caption{Catalogue of Variable Point Sources in M33 - 1. Chip numbers are as shown in fig.~\ref{fov}. $g^{\prime}$, $r^{\prime}$ and $i^{\prime}$ are the magnitudes measured on the reference image. The listed errors on these measurements are the formal uncertainties returned by DAOPHOT. $g^{\prime}flag$, $r^{\prime}flag$ and $i^{\prime}flag$ denote whether or not the object was detected as a variable source in this filter.}
\label{tab:varcat1}
\begin{tabular}{@{}rrrrrrrrrrccc@{}}
\hline
 ID & Chip & $\alpha_{2000}$\footnote{J2000, hours:minutes:seconds} & $\delta_{2000}$\footnote{J2000, degrees:minutes:seconds} & $g^{\prime}$\footnote{magnitudes}\footnote{Value is 99.999 for no data.} & $g^{\prime}err$ & $r^{\prime}$ & $r^{\prime}err$ & $i^{\prime}$ & $i^{\prime}err$ & $g^{\prime}flag$\footnote{Value is 1 if the object was detected as variable in this filter, 0 otherwise.} & $r^{\prime}flag$ & $i^{\prime}flag$ \\
\hline
10001 & 1 & 01:35:39.38 & 31:00:51.2 & 23.519 & 0.023 & 22.389 & 0.009 & 20.848 & 0.004 & 0 & 1 & 1 \\
10002 & 1 & 01:35:39.79 & 30:57:15.2 & 24.185 & 0.030 & 22.093 & 0.009 & 21.137 & 0.008 & 0 & 1 & 1 \\
10003 & 1 & 01:35:40.29 & 31:05:36.1 & 24.577 & 0.026 & 22.386 & 0.006 & 21.396 & 0.005 & 0 & 1 & 1 \\
10004 & 1 & 01:35:43.17 & 31:02:39.9 & 22.924 & 0.009 & 99.999 & 99.999 & 20.369 & 0.006& 0 & 0 & 1  \\
10005 & 1 & 01:35:44.04 & 31:05:14.0 & 99.999 & 99.999 & 23.443 & 0.021 & 21.158 & 0.005& 0 & 0 & 1  \\
10006 & 1 & 01:35:44.95 & 31:06:31.5 & 23.878 & 0.015 & 22.522 & 0.007 & 20.769 & 0.005 & 0 & 0 & 1 \\
10007 & 1 & 01:35:47.46 & 31:05:11.8 & 99.999 & 99.999 & 99.999 & 99.999 & 23.394 & 0.035 & 0 & 0 & 1 \\
10008 & 1 & 01:35:47.57 & 31:02:42.6 & 24.646 & 0.296 & 99.999 & 99.999 & 99.999 & 99.999& 1 & 0 & 0 \\
10009 & 1 & 01:35:47.64 & 31:00:19.2 & 23.927 & 0.024 & 22.568 & 0.009 & 20.820 & 0.006 & 0 & 1 & 1 \\
10010 & 1 & 01:35:48.38 & 31:06:15.0 & 21.148 & 0.005 & 20.944 & 0.005 & 20.551 & 0.004 & 1 & 0 & 0 \\
\hline
\end{tabular}
\end{minipage}
\end{table*}

\begin{table*}
\centering
\begin{minipage}{168mm}
\caption{Catalogue of Variable Point Sources in M33 - 2. The minimum and maximum magnitudes in each filter are the observed minimum and maximum magnitudes, no attempt has been made to cull out-lier points in determining these values. $g^{\prime}avg$, $r^{\prime}avg$ and $i^{\prime}avg$ refer to the flux-averaged magnitude of the light curves. $g^{\prime}rms$, $r^{\prime}rms$ and $i^{\prime}rms$ are the magnitude RMS values for the light curves, again determined without culling outliers.}
\label{tab:varcat2}
\begin{tabular}{@{}rrrrrrrrrrrrr@{}}
\hline
 ID & $g^{\prime}max$\footnote{magnitudes}\footnote{Value is 99.999 for no data.} & $g^{\prime}min$ & $g^{\prime}avg$ &  $g^{\prime}rms$ &  $r^{\prime}max$ & $r^{\prime}min$ & $r^{\prime}avg$ & $r^{\prime}rms$ & $i^{\prime}max$ & $i^{\prime}min$ & $i^{\prime}avg$ & $i^{\prime}rms$ \\
\hline
10001 & 24.134 & 23.131 & 23.596 & 0.246 & 22.567 & 21.907 & 22.282 & 0.185 & 21.011 & 20.510 & 20.786 & 0.132 \\
10002 & 25.506 & 23.962 & 24.393 & 0.366 & 22.619 & 22.018 & 22.264 & 0.198 & 21.503 & 21.020 & 21.205 & 0.146 \\
10003 & 25.889 & 24.371 & 24.772 & 0.387 & 22.841 & 22.295 & 22.591 & 0.172 & 21.699 & 21.339 & 21.539 & 0.117 \\
10004 & 23.280 & 22.764 & 22.992 & 0.141 & 99.999 & 99.999 & 99.999 & 99.999 & 20.499 & 20.325 & 20.412 & 0.051 \\
10005 & 99.999 & 99.999 & 99.999 & 99.999 & 24.423 & 22.976 & 23.472 & 0.274 & 21.317 & 20.851 & 21.074 & 0.115 \\
10006 & 24.505 & 23.446 & 23.859 & 0.253 & 23.069 & 22.091 & 22.498 & 0.232 & 21.062 & 20.569 & 20.763 & 0.135 \\
10007 & 99.999 & 99.999 & 99.999 & 99.999 & 99.999 & 99.999 & 99.999 & 99.999 & 24.384 & 22.591 & 23.142 & 0.438 \\
10008 & 27.074 & 22.169 & 24.352 & 1.095 & 99.999 & 99.999 & 99.999 & 99.999 & 99.999 & 99.999 & 99.999 & 99.999 \\
10009 & 24.796 & 23.755 & 24.098 & 0.317 & 23.480 & 22.506 & 22.845 & 0.273 & 21.429 & 20.774 & 20.996 & 0.195 \\
10010 & 21.331 & 20.951 & 21.192 & 0.079 & 21.077 & 20.810 & 20.984 & 0.063 & 20.681 & 20.495 & 20.591 & 0.043 \\
\hline
\end{tabular}
\end{minipage}
\end{table*}

\begin{table*}
\centering
\begin{minipage}{168mm}
\caption{Catalogue of Variable Point Sources in M33 - 3. ID$_{M}$ is the ID of the source in the M06 catalogue. Columns beginning with D33 are the IDs of the source in the DIRECT catalogues. Comments denote pairs of objects that are within $0\farcs3$ of one another, matches to Chandra sources (Grimm et al. 2005), or matches to H-$\alpha$ emission line sources (Calzetti et al. 1995). Rows containing matches to M06 or DIRECT, or comments have been selected to illustrate the content of the catalogue.}
\label{tab:varcat3}
\begin{tabular}{@{}rlllll@{}}
\hline
 ID & ID$_{M}$\footnote{Massey et al. (2006)} & D33AID\footnote{Mochejska et al. (2001a)} & D33BID\footnote{Mochejska et al. (2001b)} & D33ABID\footnote{Macri et al. (2001)} & Comments \\
\hline
30073 & J013440.10+305905.7 & ~$\cdots$~ & ~$\cdots$~ & ~$\cdots$~ & Close Pair 30074 $\delta=0.26$; \\
30074 & ~$\cdots$~ & ~$\cdots$~ & ~$\cdots$~ & ~$\cdots$~ & Close Pair 30073 $\delta=0.26$; \\ 
30158 & ~$\cdots$~ & ~$\cdots$~ & ~$\cdots$~ & ~$\cdots$~ & Chandra J013444.6+305535 $\delta=0.50$; \\
40584 & J013424.17+305631.0 & ~$\cdots$~ & ~$\cdots$~ & ~$\cdots$~ & Close Pair 40586 $\delta=0.13$; \\ 
40586 & ~$\cdots$~ & ~$\cdots$~ & ~$\cdots$~ & ~$\cdots$~ & Close Pair 40584 $\delta=0.13$; \\ 
130112 & J013407.23+304158.8 & ~$\cdots$~ & ~$\cdots$~ & D33J013407.2+304159.3 & ~$\cdots$~ \\
130113 & J013407.21+304635.0 & D33J013407.3+304635.5 & ~$\cdots$~ & D33J013407.3+304635.5 & ~$\cdots$~ \\
130349 & ~$\cdots$~ & D33J013408.5+304430.6 & ~$\cdots$~ & D33J013408.5+304430.6 & ~$\cdots$~ \\
130408 & J013408.72+304543.1 & D33J013408.8+304543.5 & ~$\cdots$~ & D33J013408.8+304543.5 & ~$\cdots$~ \\
130504 & ~$\cdots$~ & D33J013409.3+304238.6 & ~$\cdots$~ & D33J013409.3+304238.6 & ~$\cdots$~ \\
\hline
\end{tabular}
\end{minipage}
\end{table*}

\subsection{Comparison with Other Catalogues}

\begin{figure*}

\includegraphics[width=168mm]{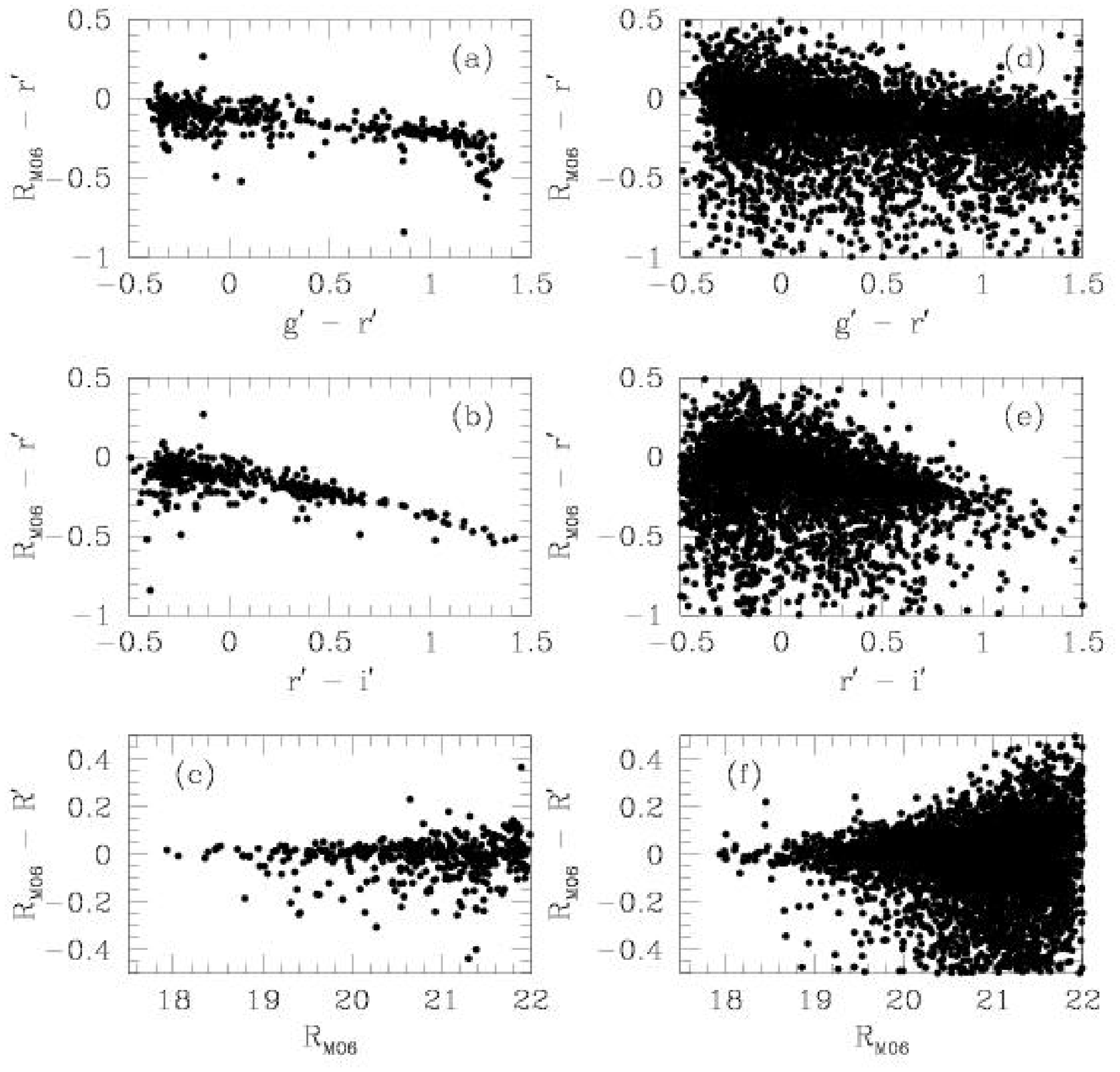}
\caption{A comparison between our photometry ($g^{\prime}$,$r^{\prime}$,$i^{\prime}$) and the photometry from M06 ($R_{M06}$) for chips 3 (a-c) and 14 (d-f). There is a definite colour dependence in the transformation from $r^{\prime}$ to $R_{M06}$, plots (a) and (d) show the dependence of this transformation on $g^{\prime}-r^{\prime}$ and plots (b) and (e) show the dependence on $r^{\prime}-i^{\prime}$. The bottom two plots (c,f) show the scatter after applying the transformation, here $R^{\prime}$ represents our transformed photometry. This transformation is preliminary, we will provide a definitive photometric calibration with the complete point source catalogue in a future publication.}
\label{fig:mascompare}
\end{figure*}

The catalogue includes matches to both the variable star catalogues from DIRECT and the point source catalogue from M06. In fig.~\ref{fig:mascompare} we compare the $r^{\prime}$ instrumental photometry from our constant stars with the $R$ photometry from M06 for chips 3 and 14. The transformation is strongly colour dependent, and varies from chip to chip. For each chip with at least 15 matches we fit a transformation of the form
\begin{equation}
R = r^{\prime} + a0 + a1(g^{\prime}-r^{\prime}) + a2(r^{\prime}-i^{\prime}). \label{eq:transformation}
\end{equation}
Table~\ref{tab:trans} lists the resulting parameters of this fit along with the standard deviation after applying this fit to stars with $R < 20$. We take the standard deviation after applying an iterative $3\sigma$ clipping to remove mismatches or stars that were resolved into different numbers of point sources in the two catalogues. We find that the scatter for bright stars is below 0.1 mag for all of the chips and below 0.03 mag for many of the chips that do not cover the centre of the galaxy. Though we do not apply these transformations to our catalogue, we provide them as a preliminary calibration.

\begin{table}
\caption{Parameters for Photometric Transformation To $R$ (M06). The parameters are defined in eq.~\ref{eq:transformation}. Only chips with more than 15 stars in common with M06 are listed. The scatter is the standard deviation after an iterative $3\sigma$ clipping for stars with $R < 20$.}
\label{tab:trans}
\begin{tabular}{@{}rrrrr@{}}
\hline
 Chip & $a_{0}$ & $a_{1}$ & $a_{2}$ & Scatter \\
\hline
 2 & -0.1216 & -0.1010 & -0.1511 & 0.041 \\
 3 & -0.1231 & -0.0630 & -0.1031 & 0.030 \\
 4 & -0.0807 & -0.1466 &  0.0118 & 0.042 \\
 5 & -0.1007 & -0.1333 &  0.0064 & 0.027 \\
 6 & -0.1302 & -0.1783 & -0.0313 & 0.085 \\
 7 & -0.1579 &  0.0310 & -0.3168 & 0.020 \\
11 & -0.1490 & -0.1165 & -0.0446 & 0.093 \\
12 & -0.0934 & -0.1041 & -0.0265 & 0.034 \\
13 & -0.0933 & -0.0777 & -0.0403 & 0.045 \\
14 & -0.0723 & -0.0836 & -0.0869 & 0.058 \\
15 & -0.1154 & -0.0845 & -0.0755 & 0.034 \\
16 & -0.1671 & -0.0904 & -0.0493 & 0.067 \\
17 & -0.1849 & -0.0507 & -0.1546 & 0.044 \\
20 & -0.1654 &  0.0620 & -0.3213 & 0.021 \\
21 & -0.1229 & -0.1205 & -0.0091 & 0.024 \\
22 & -0.1197 & -0.0618 & -0.0589 & 0.045 \\
23 & -0.0973 &  0.1068 & -0.4112 & 0.065 \\
24 & -0.0694 & -0.0629 & -0.0536 & 0.032 \\
25 & -0.0840 & -0.0911 & -0.0531 & 0.024 \\
26 & -0.1054 & -0.0805 & -0.0968 & 0.025 \\
27 & -0.2358 &  0.2651 & -0.4916 & 0.026 \\
31 & -0.1251 & -0.0793 & -0.1142 & 0.030 \\
32 & -0.1038 & -0.0490 & -0.1288 & 0.019 \\
33 & -0.0989 & -0.0985 & -0.0352 & 0.023 \\
34 & -0.0790 & -0.0857 & -0.0747 & 0.018 \\
35 & -0.1265 & -0.0373 & -0.1785 & 0.061 \\
36 & -0.0893 & -0.0803 & -0.2243 & 0.027 \\
\hline
\end{tabular}
\end{table}

We also compared our catalogue with the lists of variables found by the DIRECT project. We find matches within $1\farcs0$ to 1092 out of 1430 distinct variable sources. Of the 338 variables that we do not match with, $\sim 110$ lie on or near a chip gap or near an image artifact. An inspection of the spatial distribution of DIRECT sources that did not match to our catalogue revealed that many of these sources are located near the densest part of the image where crowding is particularly significant and thus the coordinates are uncertain. We find that when comparing to the DIRECT AB catalogue \citep{mac01} the median matching error is $0\farcs11$, while for the DIRECT A and B catalogues (Mochejska et al. 2001a, 2001b) the matching error is $0\farcs32$ and $0\farcs34$ respectively. 

We have also matched our data to a catalogue of Chandra X-Ray sources in M33 \citep{gri05}, and a catalogue of H-$\alpha$ line emission candidates \citep{cal95}. We find that out of 261 distinct Chandra sources, 60 have an optical variable counterpart, where we used a matching radius of $1\arcsec$ plus the uncertainty in the Chandra position. For the match to Calzetti et al. (2005) we find that 58 out of the 153 sources have a counterpart in our catalogue when using a matching radius of $0\farcs5$.  The IDs for these matches are provided in the comments column of the catalogue.

\section{Results}

\subsection{Colour-Magnitude Diagrams}

\begin{figure*}

\includegraphics[width=168mm]{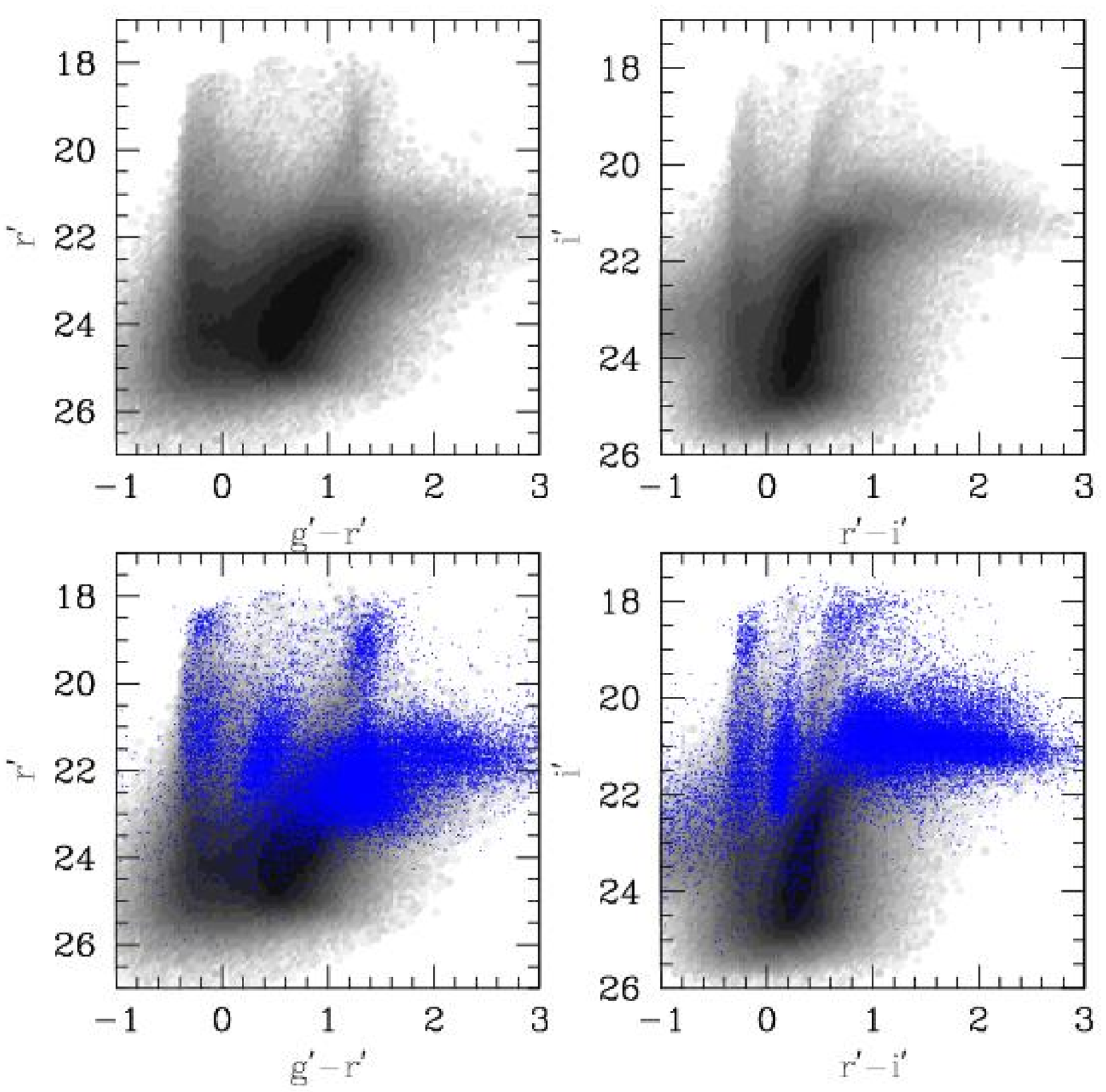}
\caption{$g^{\prime}-r^{\prime}$ vs. $r^{\prime}$ and $r^{\prime}-i^{\prime}$ vs. $i^{\prime}$ CMDs. We have adopted the Hess technique of using intensity scaling proportional to the logarithm of the number of stars located at a given colour and magnitude. In the bottom two figures we use points to show the location of all variables in our catalogue on these CMDs. The preponderance of red variables along the giant branch and the location of the Cepheid instability strip between the main sequence and giant branch are apparent.}
\label{fig:cmd}
\end{figure*}

A colour-magnitude diagram (CMD) is perhaps the simplest plot one can make that reveals information about the classes of variables present in a catalogue. In fig~\ref{fig:cmd} we present $g^{\prime}-r^{\prime}$ vs. $r^{\prime}$ and $r^{\prime}-i^{\prime}$ vs. $i^{\prime}$ CMDs with and without the variables. 

\begin{figure*}
\includegraphics[width=168mm]{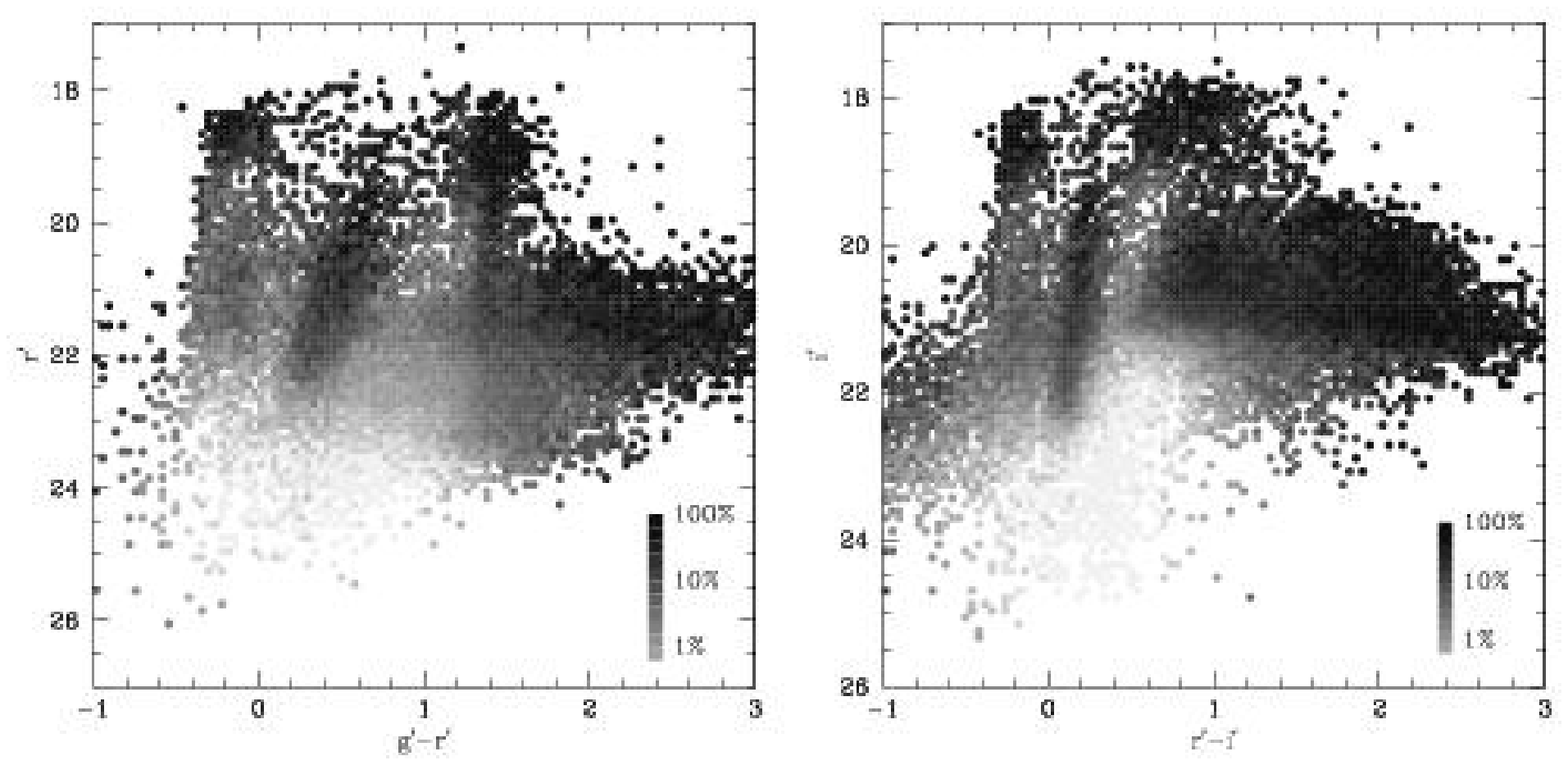}
\caption{The fraction of stars that are detected as variable on the $g^{\prime}-r^{\prime}$ vs. $r^{\prime}$ and $r^{\prime}-i^{\prime}$ vs. $i^{\prime}$ CMDs. Using these diagrams we can select classes of variables for further investigation, as done in fig.~\ref{fig:key}.}
\label{fig:frac_var}
\end{figure*}

In fig~\ref{fig:frac_var} we show the fraction of stars on the CMDs that were detected as variables. While biased, this plot allows us to identify several striking groups of variables. Two features that are immediately apparent are the Cepheid instability strip (IS) located between the main sequence and giant branches and the red long period variables (LPVs) at $g^{\prime}-r^{\prime} > 1.25, r^{\prime} > 20$. Not unexpectedly there is a sizeable group of variables along the main sequence at $-0.5 < r^{\prime}-i^{\prime} < 0, r^{\prime} > 20$, though the fraction of stars in this region that are variable does not increase dramatically. We also identify two groups of variables near the bright end of the main sequence and giant branches---at $r^{\prime}$ $\sim$ 18 to 19, and $g^{\prime}-r^{\prime}$ $\sim$ -0.2 and +1.4, respectively. These we refer to tentatively as variable blue and red supergiant stars; they may be related to the $\alpha$-Cygni type variables which are semi-regular pulsating supergiants (see for example Sterken 1996). The majority of variable candidates blueward of $g^{\prime}-r^{\prime} = -0.5$ or $r^{\prime}-i^{\prime} = -0.5$ likely have incorrect magnitudes in at least one filter since they occur predominately in, and uniformly across, a handful of chips near the centre of the galaxy where magnitude measurements and matching between filters is the most uncertain. 

\begin{figure*}

\includegraphics[width=168mm]{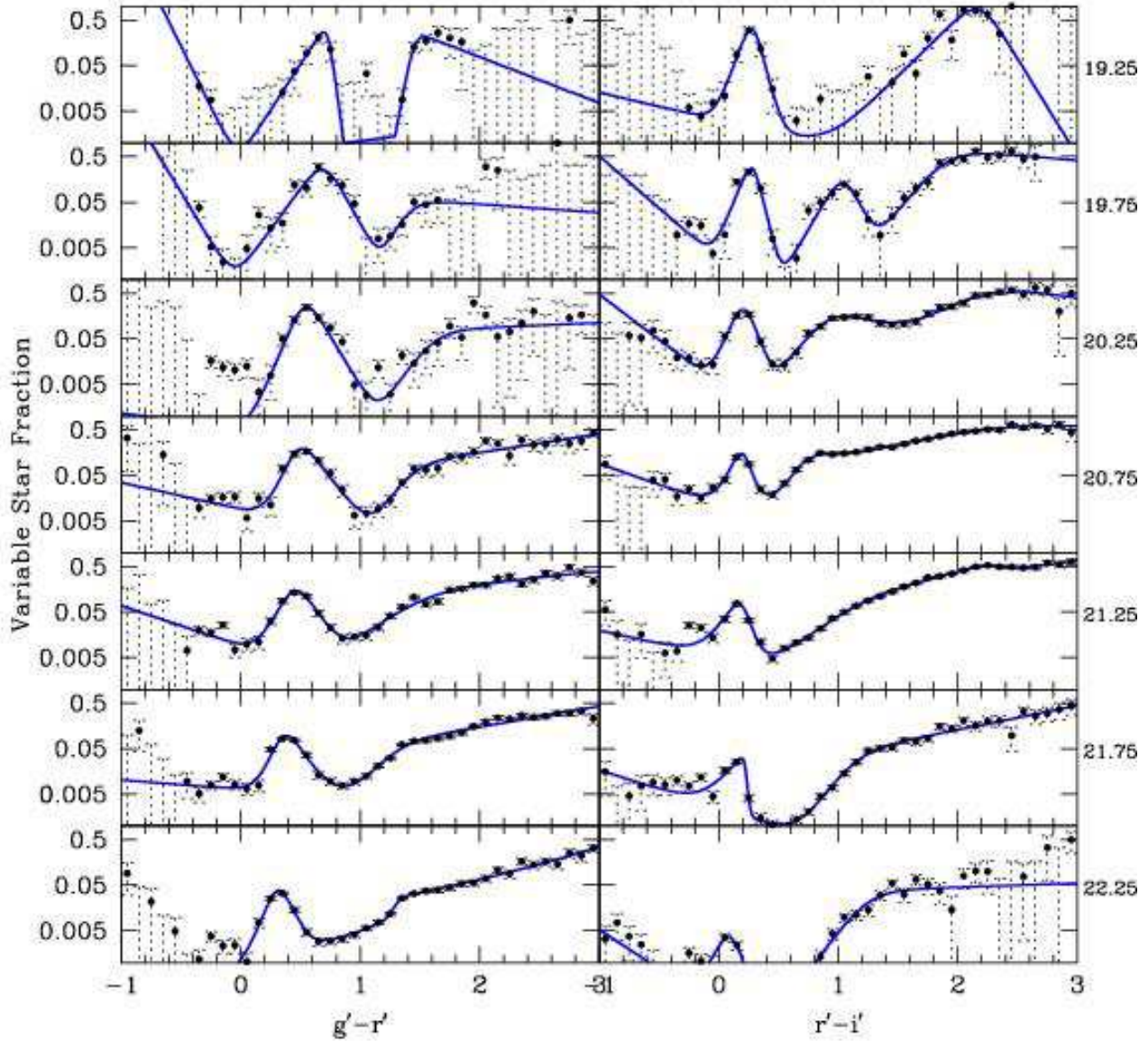}
\caption{The fraction of stars that are variable with $RMS > 0.1$ mag. as a function of colour ($g^{\prime}-r^{\prime}$ left, $r^{\prime}-i^{\prime}$ right) in magnitude bins of width 0.5 mag ranging from 19.25 to 22.25 ($r^{\prime}$ left, $i^{\prime}$ right). The error bars assume Poisson statistics. The solid lines show model fits to the observed distributions. The model is composed of a series of broken power-laws as described in \S 4.1. It is used to identify the peak of the Cepheid instability strip (between $0<g^{\prime}-r^{\prime}<1$ or $0<r^{\prime}-i^{\prime}<0.5$), its extent (which we take to be the colour at which the fraction of stars that are variable with $RMS > 0.1$ mag falls to a tenth of the peak value) and the blue edge of the long period variable population (which we take to be the location of the first power-law break redward of the Cepheid instability strip).}
\label{fig:colourdist}
\end{figure*}

To find the blue and red edges of the IS, and the blue edge of the LPV region, we plot the fraction of stars that are detected as variable with $RMS > 0.1$ mag as a function of $g^{\prime}-r^{\prime}$ and $r^{\prime}-i^{\prime}$ colours in fig~\ref{fig:colourdist}.  We determine this fraction for stars with $r^{\prime},i^{\prime}$ between $19.0$ and $22.5$ in bins of width $0.5$ mag. The location of the IS is visible in each of these distributions as a peak located between $0<g^{\prime}-r^{\prime}<1$ and $0<r^{\prime}-i^{\prime}<0.5$. Redward of the IS the fraction of stars that are variable rises until it reaches a break at around $g^{\prime}-r^{\prime} \sim 1.5$ after which the rise is slower. We fit a simple model to the observed distribution of the form
\begin{equation}
f(x) = \frac{A}{\left (\frac{x}{B} \right ) ^{C} + \left (\frac{x}{B} \right ) ^{D}} + \frac{E}{\left (\frac{x}{F} \right ) ^{G} + \left (\frac{x}{F} \right ) ^{H}} + Ix^{J},
\end{equation}
where $x$ is $\exp(g^{\prime}-r^{\prime})$ or $\exp(r^{\prime}-i^{\prime})$. The $Ix^{J}$ term is used to fit the distribution blueward of $\sim 0$, the first broken power-law is used to fit the IS and the second broken power-law is used to fit the LPV ridge. For the $r^{\prime}-i^{\prime}$ distributions between $19.5 < i^{\prime} < 21.5$ we include a third broken power-law to account for the observed bump in the LPV region, particularly noticeable for the $i^{\prime} = 19.75$ and $i^{\prime} = 20.25$ bins. Using this simple model we can define four interesting locations. The first is the peak in the IS given by $(g^{\prime}-r^{\prime})_{peak} = \ln(B)$ (similarly for $r^{\prime}-i^{\prime}$). The second and third are the blue and red edges of the IS, which we will take to be the colours at which the fraction of stars that are variable falls to a tenth its value at the peak, and which we approximate by $(g^{prime}-r^{\prime})_{ISB} = \ln(B) + \frac{1}{C}\ln(20)$ and $(g^{\prime}-r^{\prime})_{ISR} = \ln(B) + \frac{1}{D}\ln(20)$. The final location of interest is the blue edge of the LPV region which we take to be the location of the first break redward of the IS and is given by $(g^{\prime}-r^{\prime})_{LPV} = \ln(F)$.  

\begin{figure*}

\includegraphics[width=168mm]{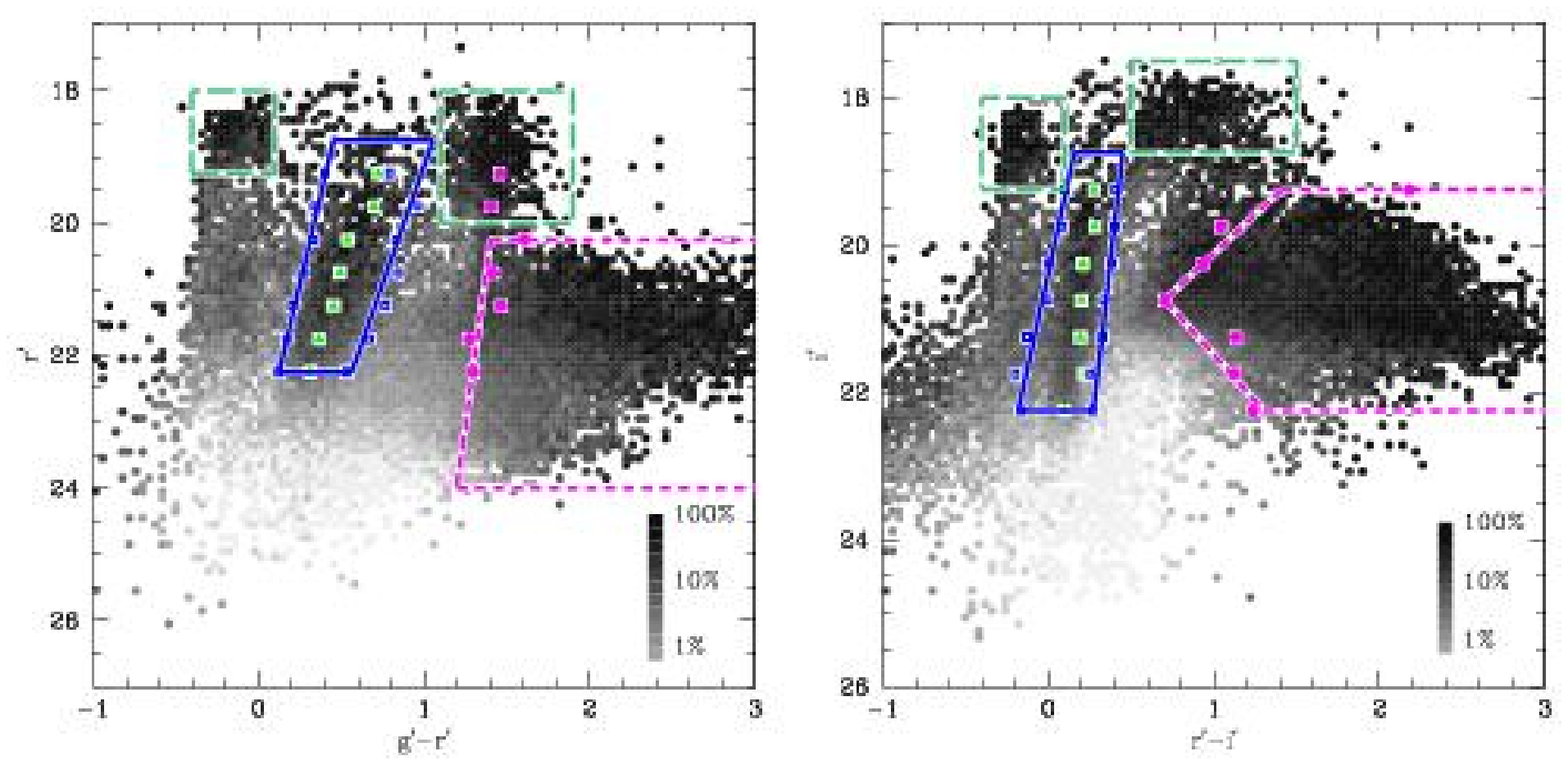}
\caption{Same as fig.~\ref{fig:frac_var}. Here we show classes of variables that we will use for further investigation. The solid lines bound the Cepheid region, the short-dashed lines bound the LPV region, and the long dashed lines bound the variable blue and red supergiants. We also plot the location of the Cepheid IS peak as found in fig.~\ref{fig:colourdist} using triangles, its blue and red edges using squares, and the blue edge of the long period variable region using circles.}
\label{fig:key}
\end{figure*}

In fig~\ref{fig:key} we plot the measured positions of the peak in the IS, its blue and red edges, and the blue edge of the LPV region. We fit a line through the blue and red edges of the Cepheid IS and the blue edge of the LPV region and find that these locations are given approximately by the relations
\begin{eqnarray}
r^{\prime}_{ISB} &=& -10.64 (\pm 0.29) (g^{\prime}-r^{\prime}) + 23.58 (\pm 0.86) \\
r^{\prime}_{ISR} &=& -7.13 (\pm 0.15) (g^{\prime}-r^{\prime}) + 26.27 (\pm 0.71) \\
r^{\prime}_{LPV} &=& -18.32 (\pm 0.57) (g^{\prime}-r^{\prime}) + 45.93 (\pm 1.55) \\
i^{\prime}_{ISB} &=& -10.36 (\pm 0.30) (r^{\prime}-i^{\prime}) + 20.38 (\pm 0.84) \\
i^{\prime}_{ISR} &=& -19.94 (\pm 0.89) (r^{\prime}-i^{\prime}) + 27.72 (\pm 1.54) \\
i^{\prime}_{LPV} &=& -2.12 (\pm 0.01) (r^{\prime}-i^{\prime}) \nonumber \\
                 & & \mbox{} + 22.23 (\pm 0.11) \hspace{0.6in} i^{\prime} < 20.75 \\
i^{\prime}_{LPV} &=& 2.51 (\pm 0.02) (r^{\prime}-i^{\prime}) \nonumber \\
                 & & \mbox{} + 18.97 (\pm 0.17) \hspace{0.6in} i^{\prime} > 20.75
\end{eqnarray}
Using these relations we show the approximate locations of IS and LPV stars in fig~\ref{fig:key}. We also identify regions around the red and blue supergiant branches that have an increased fraction of variability.

\subsection{Light Curves}

\begin{figure*}

\includegraphics[width=168mm]{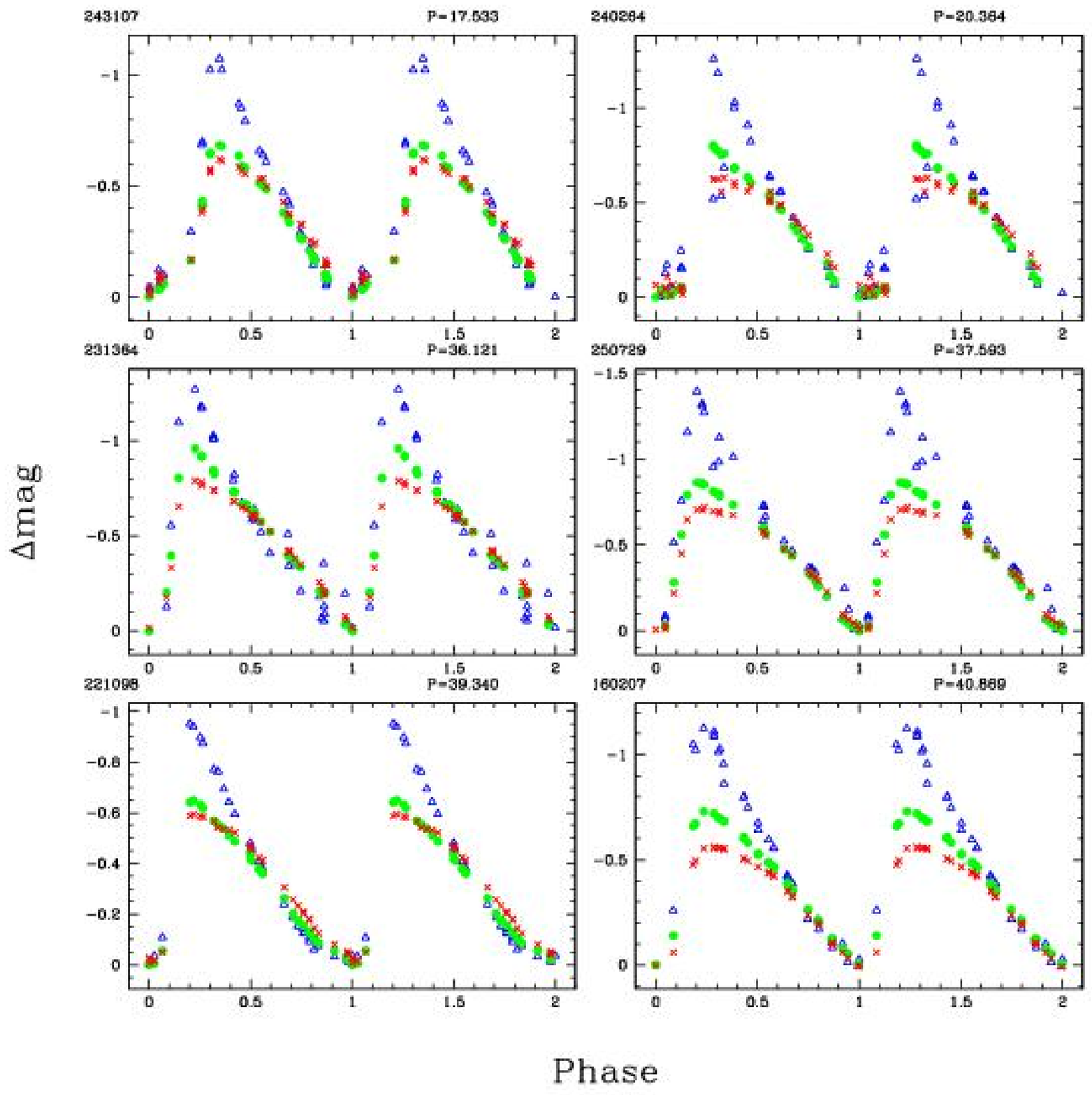}
\caption{Example phased differential magnitude light curves for 6 Cepheid variables. We use open triangles for the $g^{\prime}$ observations, filled circles for $r^{\prime}$, and Xs for $i^{\prime}$. The variables are sorted by increasing period and the light curves have been shifted so that the faintest observation in each filter is at magnitude zero. This scale makes colour variations as well as luminosity variations apparent. We also provide the ID and period (in days) of each light curve. The periods were calculated using the \citet{sch96} algorithm.}
\label{fig:Cepheidlc}
\end{figure*}

\begin{figure*}

\includegraphics[width=168mm]{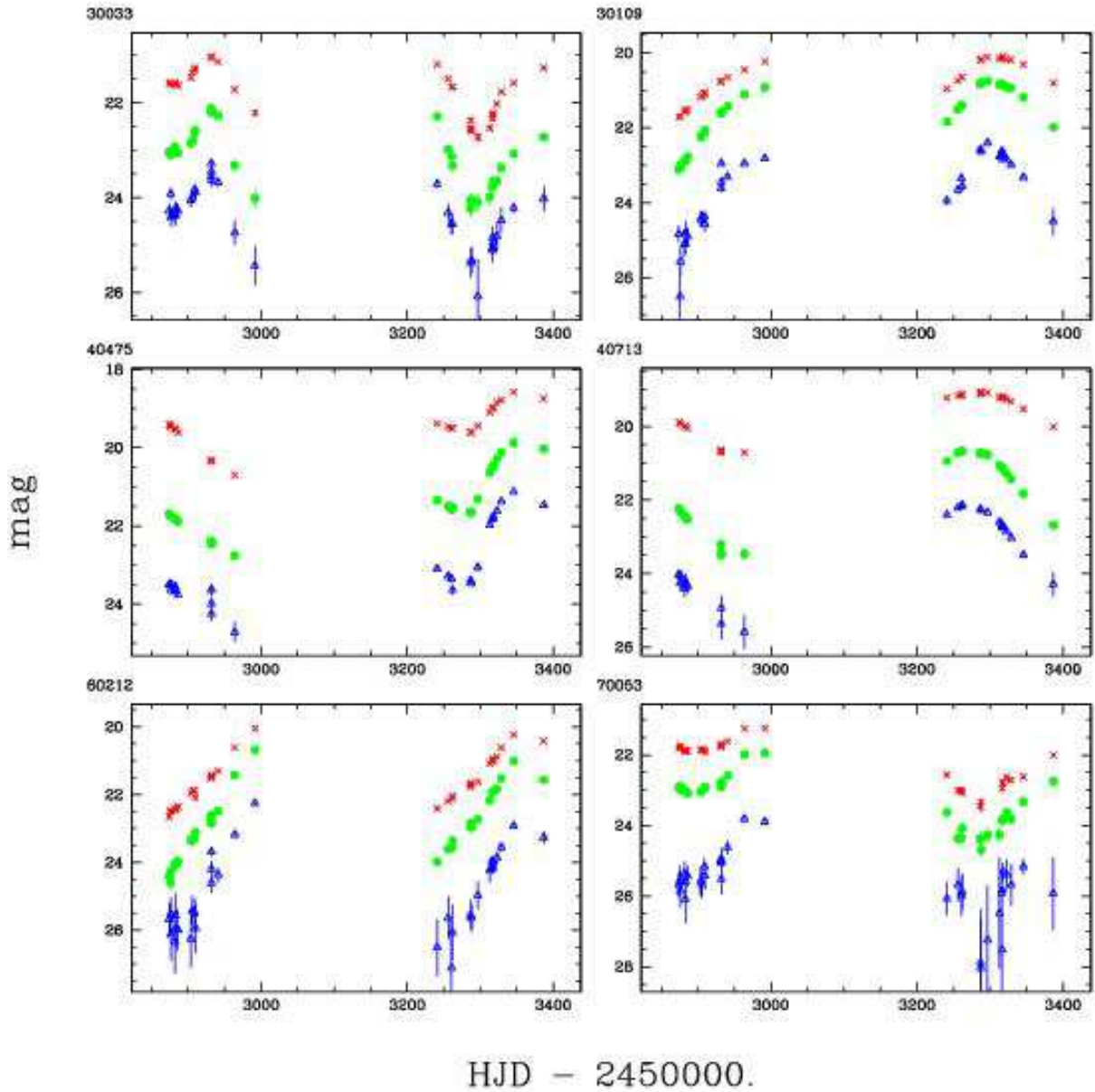}
\caption{Example light curves for 6 long period variables (LPVs). We use open triangles for the $g^{\prime}$ observations, filled circles for $r^{\prime}$, and Xs for $i^{\prime}$.}
\label{fig:LPVlc}
\end{figure*}

\begin{figure*}

\includegraphics[width=168mm]{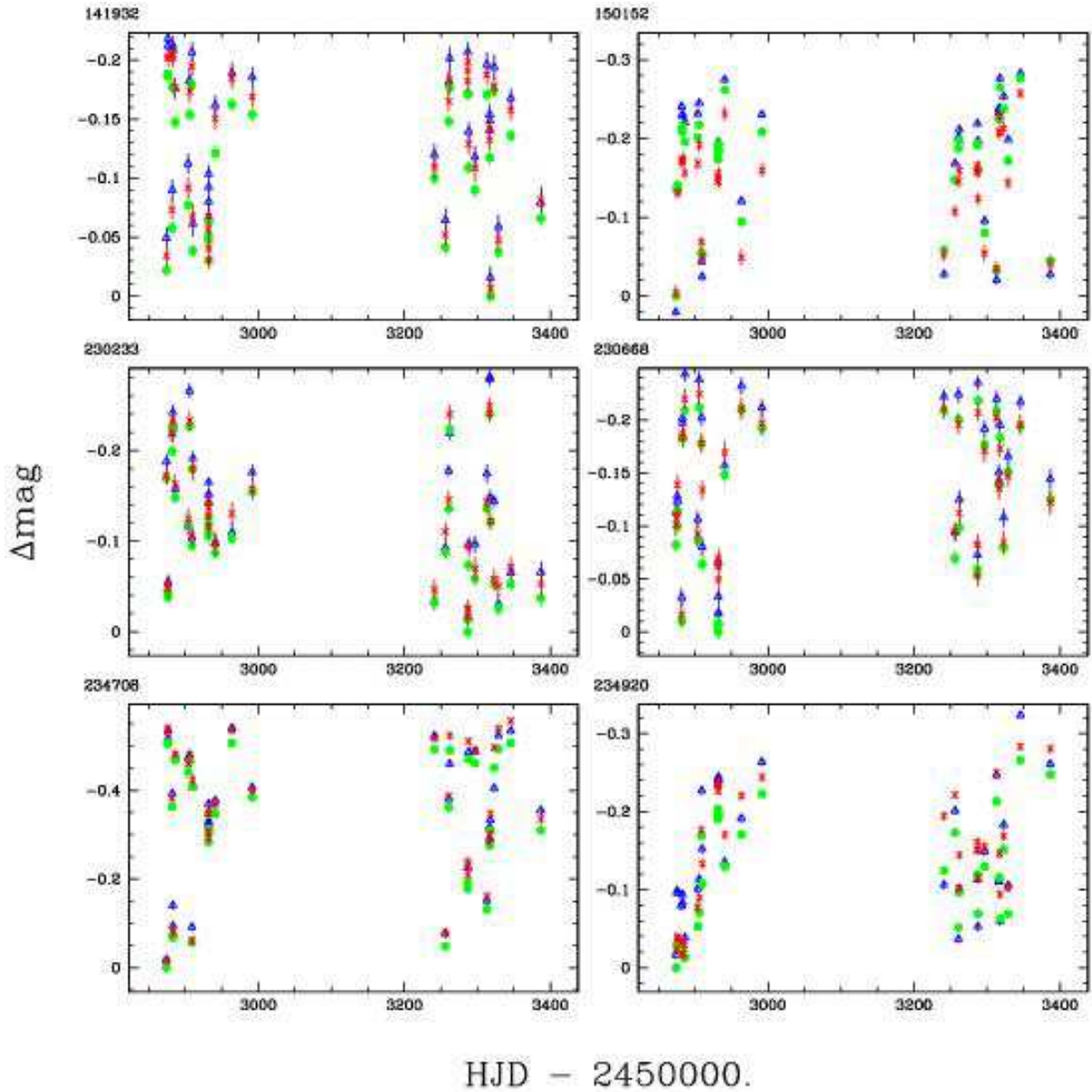}
\caption{Example differential magnitude light curves for 6 variable blue supergiants. We use open triangles for the $g^{\prime}$ observations, filled circles for $r^{\prime}$, and Xs for $i^{\prime}$. The light curves have been shifted so that the faintest observation in each filter is at magnitude zero. This scale makes colour variations as well as luminosity variations apparent.}
\label{fig:ACygBluelc}
\end{figure*}

\begin{figure*}

\includegraphics[width=168mm]{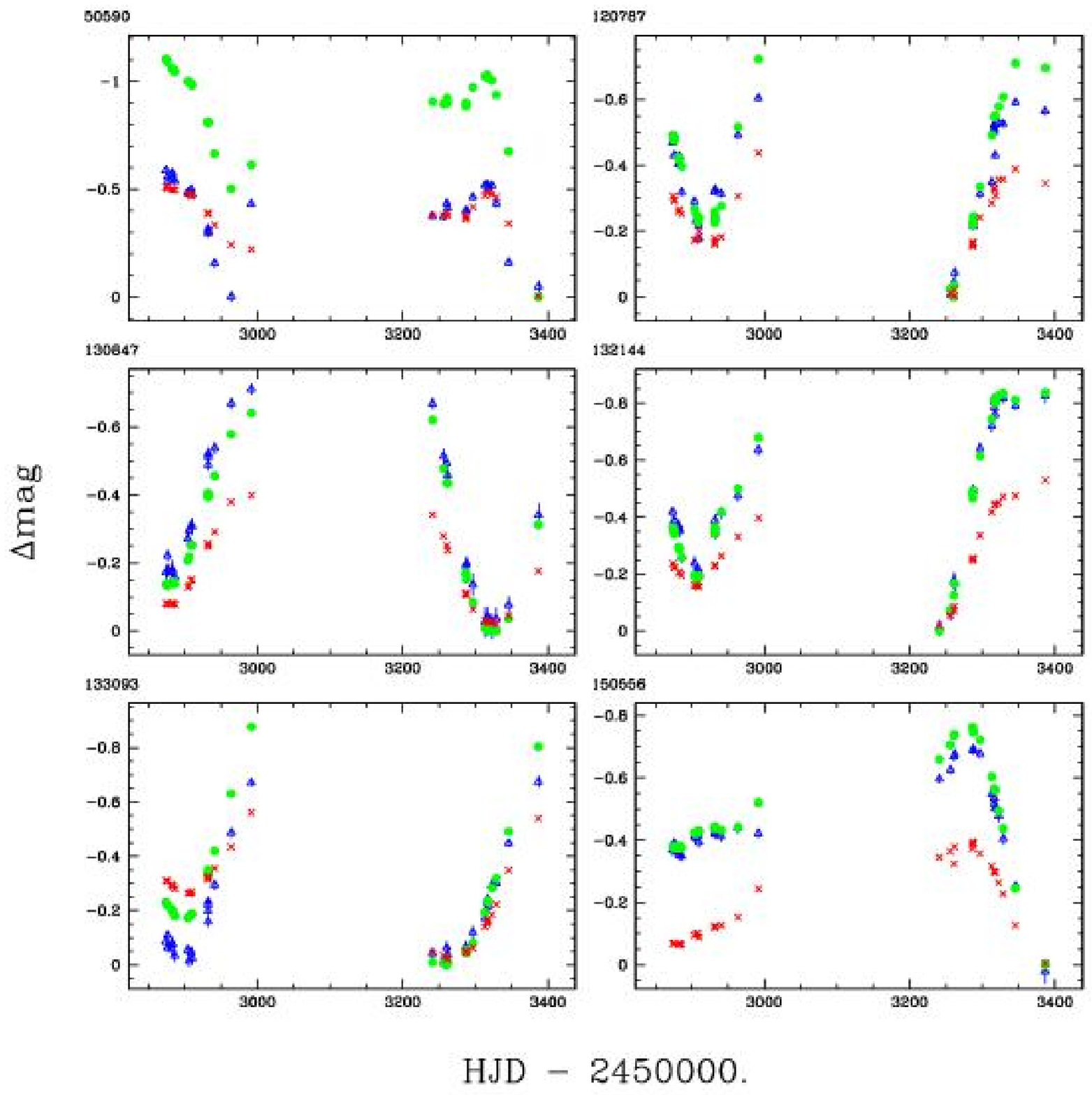}
\caption{Same as fig.~\ref{fig:ACygBluelc}, here we show variable red supergiants.}
\label{fig:ACygRedlc}
\end{figure*}

As a demonstration of our photometry we plot the $g^{\prime},r^{\prime},i^{\prime}$ light curves for several example variables from each of the four regions identified in fig~\ref{fig:key}. In fig~\ref{fig:Cepheidlc} we plot phased differential magnitude light curves for several Cepheids sorted by increasing period. We have set the faintest point in each light curve to have magnitude zero so that variations in colour as well as luminosity are apparent. In fig~\ref{fig:LPVlc} we plot the light curves for several LPVs. We do not phase these light curves or re-scale them. In fig~\ref{fig:ACygBluelc} we plot differential magnitude light curves for several variable blue supergiant candidates, and in fig~\ref{fig:ACygRedlc} we plot differential magnitude light curves for several variable red supergiant candidates. The variable blue supergiants show variations on substantially shorter time-scales than the red variables, they also tend to have lower amplitudes than the red ones.

\subsection{Spatial Distributions}

\begin{figure*}

\includegraphics[width=168mm]{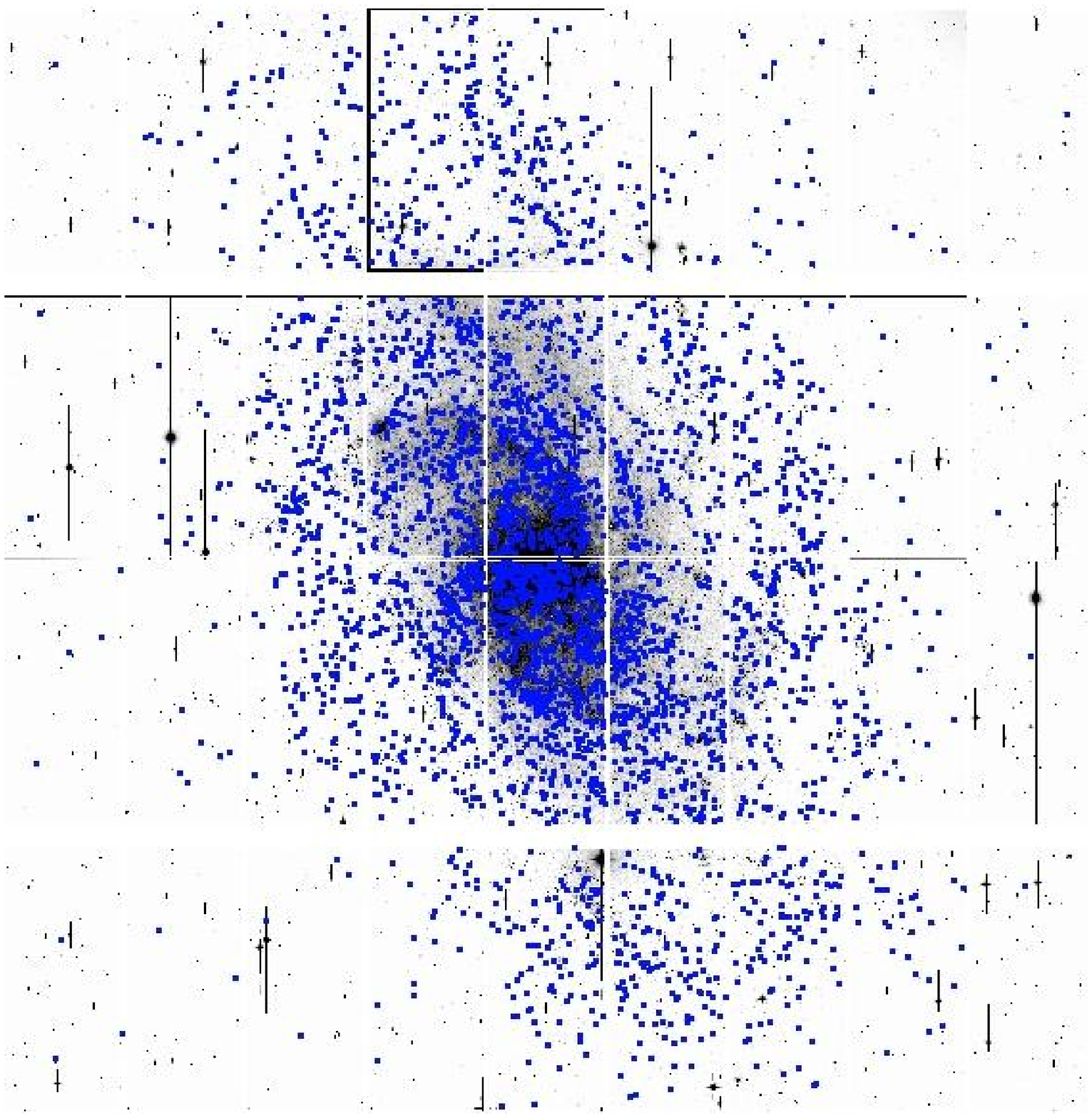}
\caption{The location of Cepheid variable stars are plotted on top of a $g^{\prime}$ image of M33. The Cepheids appear to show some correlation with the spiral arms.}
\label{fig:spatialdist_Ceph}
\end{figure*}

\begin{figure*}

\includegraphics[width=168mm]{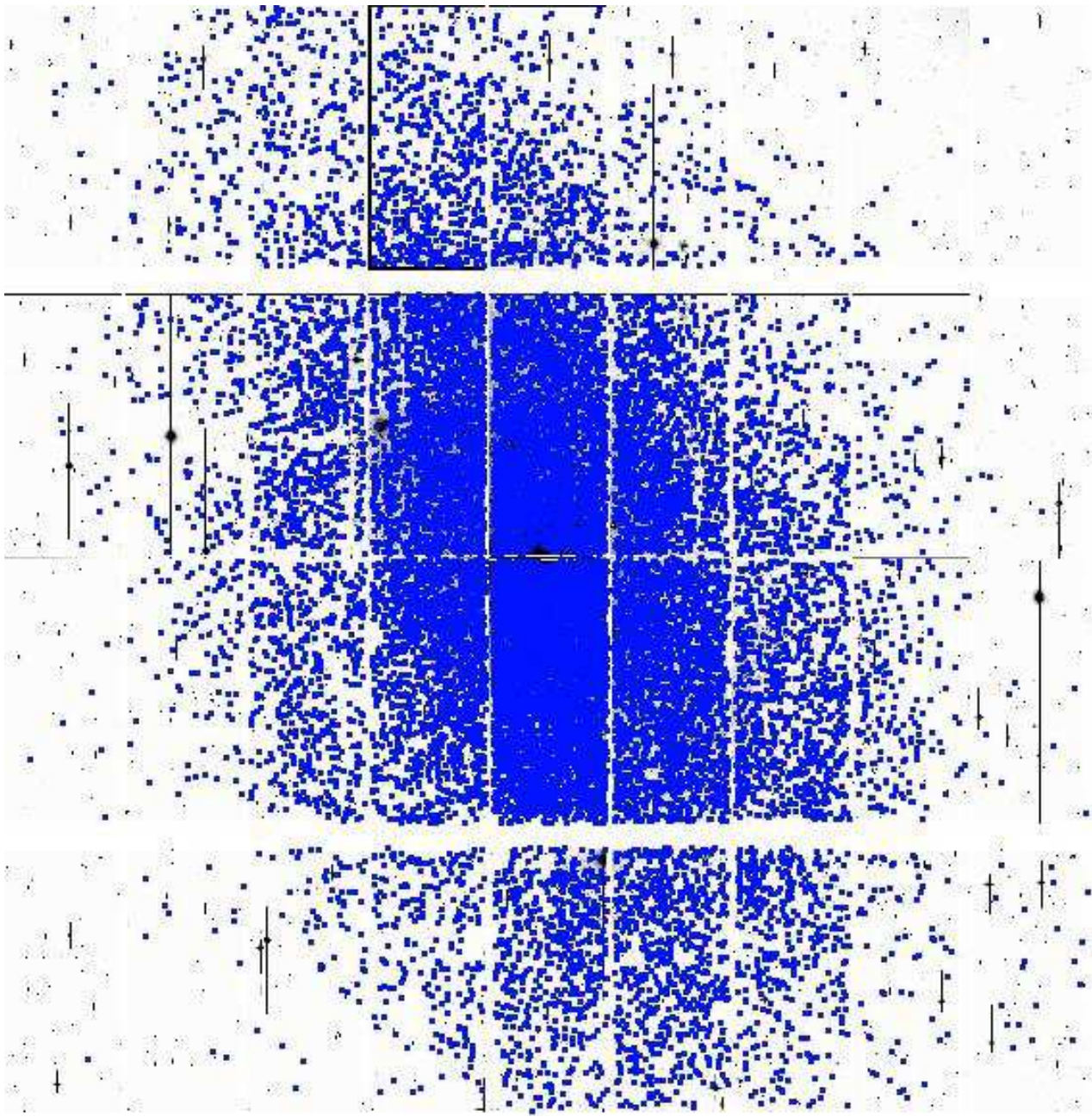}
\caption{Same as fig.~\ref{fig:spatialdist_Ceph}, here we show LPVs. The LPVs do not appear to be correlated with the spiral arms.}
\label{fig:spatialdist_LPV}
\end{figure*}

\begin{figure*}

\includegraphics[width=168mm]{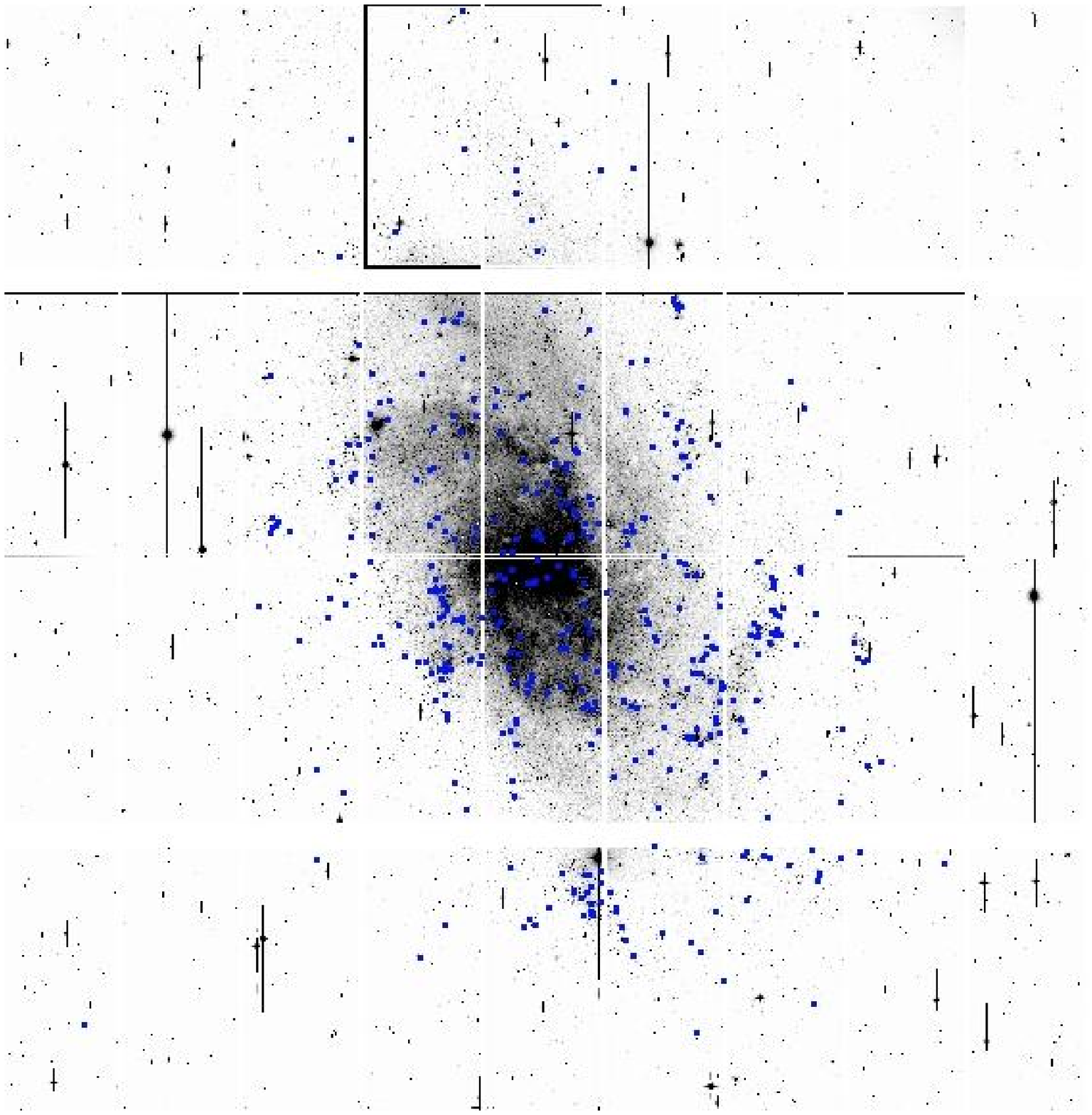}
\caption{Same as fig.~\ref{fig:spatialdist_Ceph}, here we show variable blue supergiant stars which do appear to be correlated with the spiral arms.}
\label{fig:spatialdist_VBSG}
\end{figure*}

\begin{figure*}

\includegraphics[width=168mm]{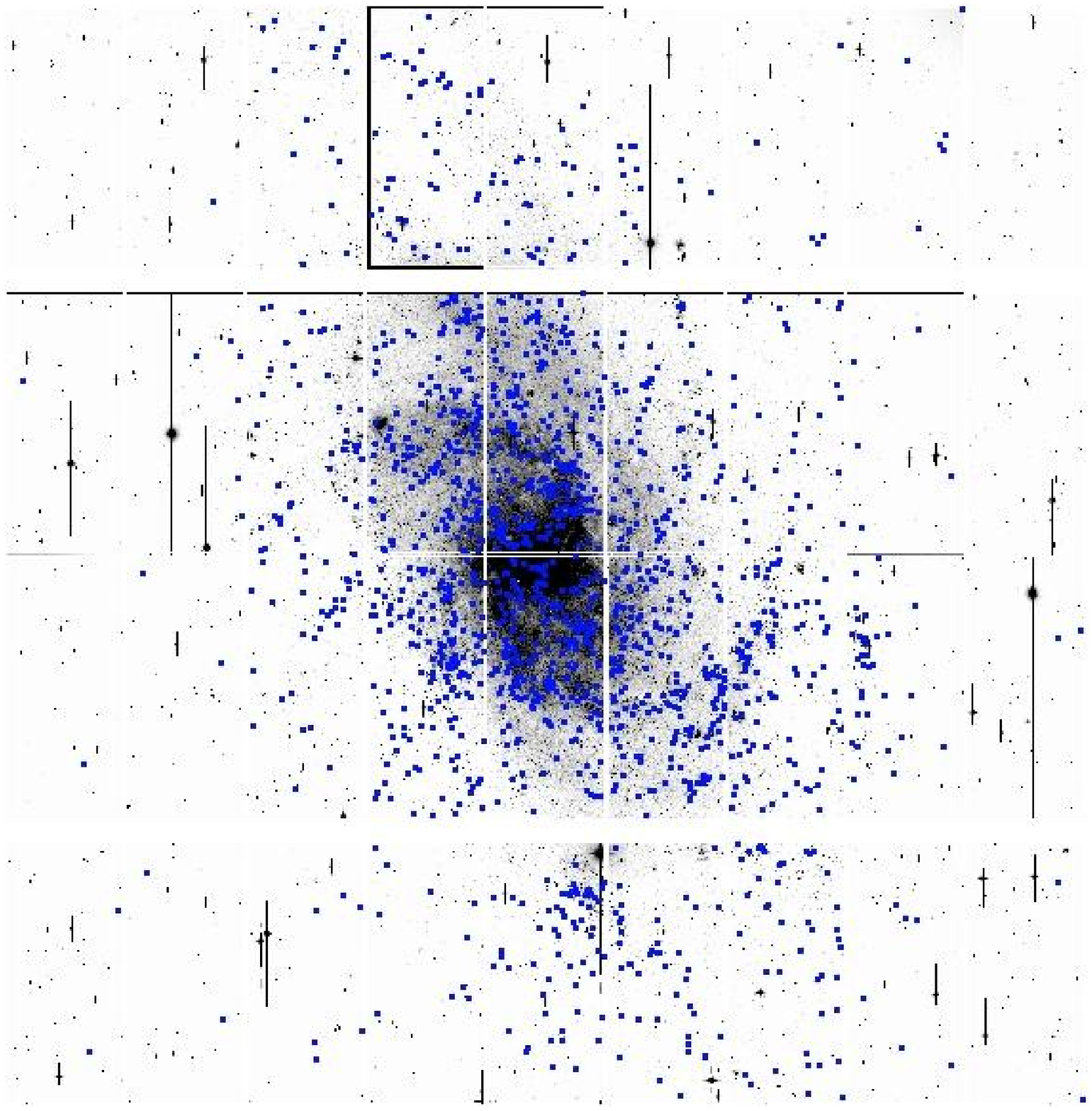}
\caption{Same as fig.~\ref{fig:spatialdist_Ceph}, here we show variable red supergiant stars which do appear to be correlated with the spiral arms.}
\label{fig:spatialdist_VRSG}
\end{figure*}

In figs.~\ref{fig:spatialdist_Ceph}-\ref{fig:spatialdist_VRSG} we show the spatial distributions of the variables identified in fig.~\ref{fig:key}. As expected the long period variables appear to trace the diffuse disk population, while the other classes of variables show a correlation with the spiral arms. Note that the variable red supergiant stars also appear to be correlated with the spiral arms, this suggests that many of these variables are in fact red supergiant stars in M33 rather than foreground giants. 

\subsection{Period-Magnitude Relation}

\begin{figure*}

\includegraphics[width=168mm]{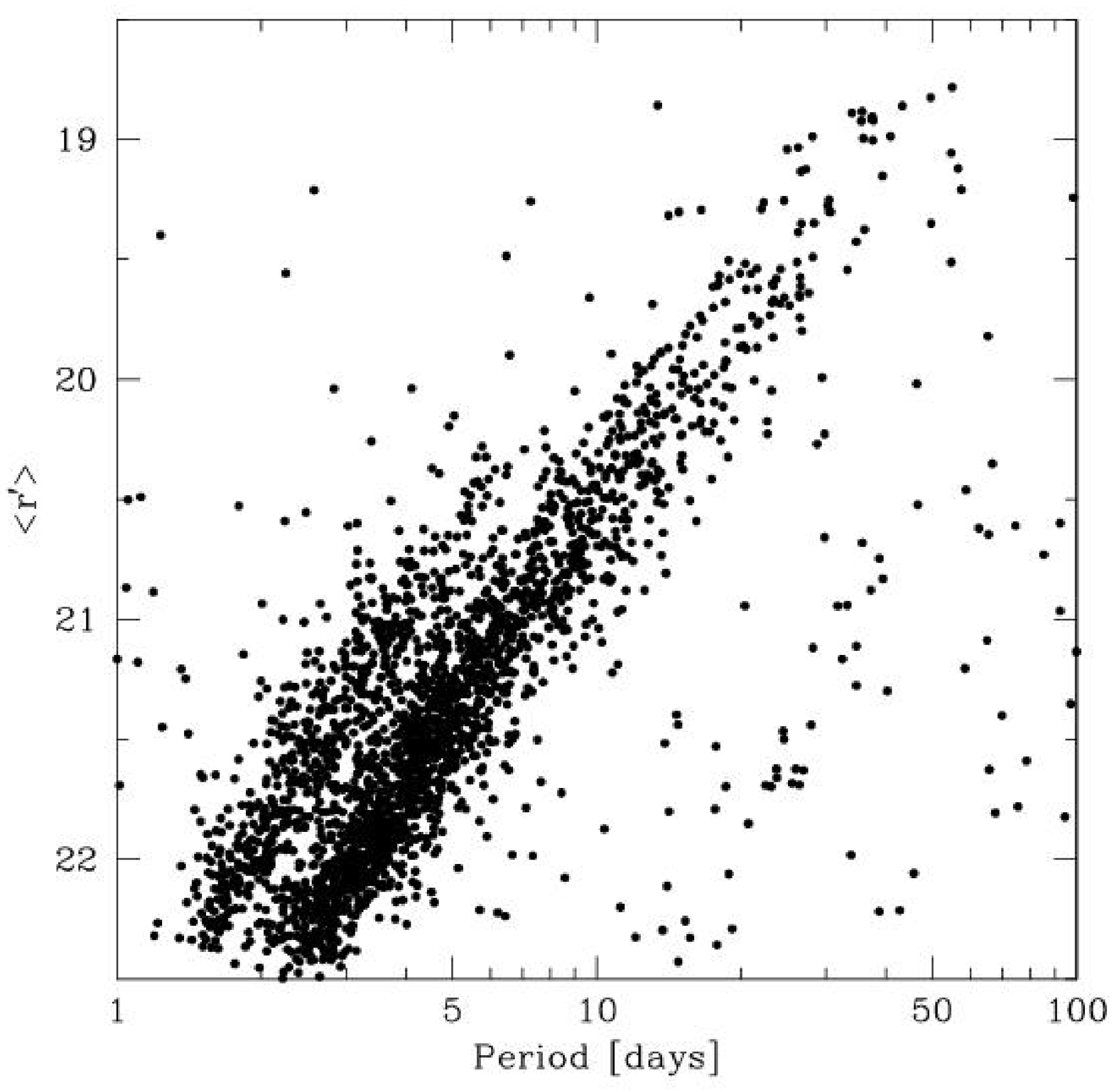}
\caption{The period-magnitude relation for $\sim 2500$ variables located in the Cepheid IS that have pulsation-like light curves. The magnitude plotted is the mean $r^{\prime}$ magnitude of the variable determined by fitting a fourier series to the light curve. No effort has been made to correct for extinction. Cepheids pulsating in the fundamental and first overtone modes are clearly visible on this diagram.}
\label{fig:Cepheid_PL}
\end{figure*}

In fig.~\ref{fig:Cepheid_PL} we show a preliminary period-magnitude diagram for the variables that lie in the Cepheid IS and have pulsation-like light curves. The magnitude plotted is the mean $r^{\prime}$ magnitude determined from fitting a Fourier series to the light curves. We show this diagram to illustrate the quality and abundance of data available for Cepheids. We will conduct an analysis of the Cepheid population of M33 in a future contribution.

\section{Summary}

We have identified more than 36,000 variable sources in M33 and are releasing photometry for all of them with multi-filter data for many of them.  This makes M33 now comparable to either of the Magellanic cloud galaxies in terms of the sheer number of variable stars that are known. These data already make possible a number of interesting projects. Studies of the period-luminosity-colour relation for Cepheids in M33, and a test of its dependence on metallicity can be done with these data. This will be done when the photometry is properly calibrated. Investigations of the $RMS$ or amplitude distributions of variables can be done using this relatively uniform data set. Studies of the eclipsing binary and ellipsoidal variable populations could be undertaken. And, with continued observations of this galaxy, it will be possible to study the LPV population for a galaxy beyond the Magellanic clouds. 

\section*{Acknowledgements}
We gratefully acknowledge the CFHT QSO observers who made this project possible, the Hawaiian people for allowing us to use a sacred mountain, G.~Pojmanski for his excellent ``lc'' program and J.~Devor for his period-finding code. JDH is funded by a National Science Foundation Graduate Student Research Fellowship. This research has made use of the SIMBAD database,
operated at CDS, Strasbourg, France.

\end{document}